# Simulation of a mathematical model of tumoral growth using finite differences

**Jesika Maganin**
Mestre em Matemática Aplicada, pelo PGMAC/UEL
Universidade Estadual de Londrina/Departamento de Matemática
Rodovia Celso Garcia Cid, PR-445, Km 380 - Campus Universitário, PR, 86057-970
E-mail: jesikamaganin@hotmail.com

**Neyva Maria Lopes Romeiro**
Doutora em Engenharia Civil/Modelagem Matemática, pela COPPE/UFRJ/RJ
Universidade Estadual de Londrina/Departamento de Matemática
Rodovia Celso Garcia Cid, PR-445, Km 380 - Campus Universitário, PR, 86057-970
E-mail: nromeiro@uel.br

**Eliandro Rodrigues Cirilo**
Doutor em Matemática Aplicada, pela USP/São Carlos/SP
Universidade Estadual de Londrina/Departamento de Matemática
Rodovia Celso Garcia Cid, PR-445, Km 380 - Campus Universitário, PR, 86057-970
E-mail: ercirilo@uel.br

**Paulo Laerte Natti**
Doutor em Física, pela USP/SP
Universidade Estadual de Londrina/Departamento de Matemática
Rodovia Celso Garcia Cid, PR-445, Km 380 - Campus Universitário, PR, 86057-970
E-mail: plnatti@uel.br

**ABSTRACT**

The work presents a study of the non-linear mathematical model of tumor growth, proposed by Kolev and Zubik-Kowal (2011). The model is described by a system composed of four partial differential equations that represent the evolution of the density of cancer cells, density of the extracellular matrix (ECM), concentration of matrix-degrading enzyme (MDE) and concentration of tissue metalloproteinase inhibitors. For numerical simulations, the finite difference method is used, in which the temporal terms of the equations are discretized using a two-stage method. In spatial terms, finite central differences are used. A study of numerical convergence for the proposed scheme is presented, using analytical solutions manufactured in a rectangular geometry. Finally, simulations of the tumor growth model are performed, using a non-regular mesh that represents the geometry of a female breast. To simulate the model in non-regular geometry, the technique used is to approximate the contour of the physical domain by mesh segments. The simulations showed that the model has important characteristics of the interactions between tumor cells and the surrounding tissue.

**Keywords:** mesh, geometry of a female breast, extracelular, cancer.

# 1 INTRODUCTION

Cancer is considered a worldwide public health problem. For the year 2030, the World Health Organization (WHO) estimates that there will be 27 million new cases, 17 million cancer deaths and 75 million people living with the disease. Therefore, understanding the mechanisms that work is increasingly important for the prevention and treatment of the disease (INCA, 2019).

The human body is made up of trillions of living cells. Normal cells grow, multiply through a process called cell division, and die in an orderly fashion. Under normal conditions, this process is controlled and is responsible for the formation, growth and regeneration of healthy body tissues. On the other hand, there are situations, such as DNA mutations that can disrupt processes, in which these cells undergo a change and lose the ability to limit and control their growth, and then multiply very quickly and without any control. The uncontrolled growth and proliferation of cells are called cancer, a set of more than one hundred diseases that leads to the formation of abnormal tissue, determining the formation of the tumor (ONCOGUIA, 2017).

According to Byrne (1999), initially solid tumors are avascular, that is, they do not have their blood supply, they rely on the diffusion of nearby vessels to supply oxygen, nutrients and to remove waste. As the tumor grows, the demand for nutrients increases, but since the flow of nutrients is small to supply the entire mass of cells, the size of the growth of the tumor becomes limited. Growth can be resumed only if the tumor becomes vascularized, that is if it is permeated by a network of capillaries. In this phase, tumor fragments that invade the blood supply are transported to other parts of the body where, if conditions are favorable, they establish secondary tumors or metastases.

To make the transition from avascular to vascular growth, the tumor undergoes a process known as angiogenesis.

Local invasion and the development of metastases are directly associated with the extracellular matrix (ECM), which is the environment in the space between cells, in which it offers adequate conditions for their growth and differentiation. ECM serves as a support, where cells can move, and it also assists the attachment of cells to tissues. Its structure consists of fibers, proteins and collagen.

The degradation of ECM becomes essential for the growth of malignant tumors, invasion, metastasis and angiogenesis. This degradation occurs through the action of matrix-degrading enzymes (MDEs), such as metalloproteinases (MMPs), which work by disorganizing the matrix, through processes that alter cell-cell and cell-matrix interactions (PEREIRA *et al*, 2005). In contrast, MMPs tissue inhibitors (TIMPs) have the main function of inhibiting enzymes, which can neutralize EDMs. Changes in homeostasis between MMPs and TIMPs have been identified in diseases associated with uncontrolled ECM renewal, such as cancer and cardiovascular diseases (AZAMBUJA *et al*, 2008).

According to Rodrigues, Pinho and Mancera (2012), due to the complexity of cancer, the construction of mathe-matical models of the disease remains a major challenge. On the other hand, it is through the development and evolution of mathematical models that describe different aspects of tumor growth and the application of computational techniques for simulation, that some characteristics and details of tumor evolution can be described, as well as effectively used in clinical laboratories.

To detect tumor progression in a relatively short time without the enormous cost of laboratory experiments, several models have been proposed in the literature.

The mathematical models that describe the invasion of cell tissue by cancer cells using systems of partial differential equations can be found in the literature, see (KOLEV; ZUBIK-KOWAL, 2011; ANDERSON et al, 2000; CHAPLAIN; LOLAS, 2005; GATENBY, GAWLINSKI, 1996).

The model used in this work was presented by Kolev and Zubik-Kowal (2011), described by a system of partial differential equations, with four variables involved in the tumor cell invasion process, resulting in the description and evolution of the density of cancer cells, the density of the extracellular matrix (ECM), the concentration of matrix-degrading enzymes (EDM) and tissue inhibitors of metalloproteinases (TIMP).

In this work, the finite difference method is used to discretize the Kolev and Zubik-Kowal (2011) model, in which the temporal terms are discretized by a two-stage method and the spatial terms, central differences. As for the nonlinear terms involved in the model, they are linearized using the Taylor series expansion.

## 2 MATHEMATICAL MODEL

The mathematical model developed in Kolev and Zubik-Kowal (2011), describes the growth of solid tumors in the avascular stage, intending to analyze the interactions between the tumor and the surrounding tissue. The model presents four partial differential equations with the following variables: density of tumor cells, the density of ECM, concentration of EDM and concentration of endogenous inhibitor TIMP, denoted by $n, f, m$ and $u$, respectively.

In this work, results will be exposed considering the two-dimensional evolution in space $(x, y)$ and time $t$. The model presented by the variables $n(x, y, t), f(x, y, t), m(x, y, t)$ and $u(x, y, t)$, is given by

$$\frac{\partial n}{\partial t} = \underbrace{d_n \nabla^2 n}_{\text{diffusion}} - \underbrace{\gamma \nabla \cdot (n \nabla f)}_{\text{haptotaxis}} + \underbrace{\mu_1 n(1 - n - f)}_{\text{proliferation}}, \qquad (1)$$

$$\frac{\partial f}{\partial t} = - \underbrace{\eta m f}_{\text{degradation}} + \underbrace{\mu_2 f(1 - n - f)}_{\text{renovation}}, \qquad (2)$$

$$\underbrace{\frac{\partial m}{\partial t} = d_m \nabla^2 m}_{\text{diffusion}} + \underbrace{\alpha n}_{\text{production}} - \underbrace{\theta um}_{\text{neutralization}} - \underbrace{\beta m}_{\text{decay}}, \qquad (3)$$

$$\frac{\partial u}{\partial t} = \underbrace{d_u \nabla^2 u}_{\text{diffusion}} + \underbrace{\xi f}_{\text{inhibits production}} - \underbrace{\theta um}_{\text{neutralization}} - \underbrace{\rho u}_{\text{decay}}. \qquad (4)$$

Furthermore, the model establishes that the migration of tumor cells creates spatial gradients that direct the migration of invasive cells by a mechanism called haptotaxis, represented by γ. The constants $d_n$, $d_m$ and $d_u$ are the diffusion constants of the density of cancer cells, EDM and inhibitor, respectively. The rate of tumor cell proliferation and the growth rate of ECM are represented by $\mu_1$ and $\mu_2$, while η, α, θ, β, ξ and ρ are positive constants.

## 3 NUMERICAL MODEL

The discretization of the differential equations in terms of the model equations (1)-(4), is obtained using the finite difference method (FDM), which consists of replacing the derivatives present in the differential equations with finite difference approximations, in which the approximation formula is obtained from the Taylor series expansion of the derived function.

To discretize the temporal terms a two-stage method is used, introducing an intermediate time level between the levels *k* and *k* + 1 (DONEA *et al.*, 2000). The two-stage technique results in an explicit and implicit stage. The first and second-order spatial terms are discretized using central finite differences.

Considering the discretizations at the time level *k* at a point (*i*, *j*), there is the explicit stage, described by

$$n|_{i,j}^{k+1/2} = n|_{i,j}^k + \frac{\Delta t}{2}\left(d_n \Delta n|_{i,j}^k - \gamma \nabla(n|_{i,j}^k \nabla f|_{i,j}^k) + \mu_1 n|_{i,j}^k (1 - n|_{i,j}^k - f|_{i,j}^k)\right),$$

$$f|_{i,j}^{k+1/2} = f|_{i,j}^k + \frac{\Delta t}{2}\left(-\eta m|_{i,j}^k f|_{i,j}^k + \mu_2 f|_{i,j}^k (1 - n|_{i,j}^k - f|_{i,j}^k)\right),$$

$$m|_{i,j}^{k+1/2} = m|_{i,j}^k + \frac{\Delta t}{2}\left(d_m \Delta m|_{i,j}^k + \alpha n|_{i,j}^k - \theta u|_{i,j}^k m|_{i,j}^k - \beta m|_{i,j}^k\right),$$

$$n|_{i,j}^{k+1/2} = n|_{i,j}^k + \frac{\Delta t}{2}\left(d_u \Delta n|_{i,j}^k + \xi f|_{i,j}^k - \theta u|_{i,j}^k m|_{i,j}^k - \rho u|_{i,j}^k\right).$$

$$u|_{i,j}^{k+1/2} = u|_{i,j}^k + \frac{\Delta t}{2}\left(d_u \Delta u|_{i,j}^k + \xi f|_{i,j}^k - \theta u|_{i,j}^k m|_{i,j}^k - \rho u|_{i,j}^k\right).$$

The scheme resulting from the discretization of the *n* and *f* equations, respectively, results in an implicit system at the time level $k + 1$ with quadratic terms. To avoid the need to solve the nonlinear system of equations, in each step of time, a numerical technique was applied by expanding the Taylor (SHEU; LIN, 2004) series in which the terms of the system are linearized.

Considering the time level *k* +1 at a point (*i*, *j*), there is the implicit stage of the method with the linearized terms, given by,

$$n|_{i,j}^{k+1} = n|_{i,j}^{k+1/2} + \frac{\Delta t}{2}(d_n \Delta n|_{i,j}^{k+1} - \gamma \nabla(n|_{i,j}^{k+1} \nabla f|_{i,j}^{k+1})$$
$$+ \frac{\Delta t}{2} \mu_1 \left( n|_{i,j}^{k+1}(1 - f|_{i,j}^{k+1}) - 2n|_{i,j}^{k+1/2} n|_{i,j}^{k+1} + n^2|_{i,j}^{k+1/2} \right),$$

$$m|_{i,j}^{k+1} = m|_{i,j}^{k+1/2} + \frac{\Delta t}{2}(d_m \Delta m|_{i,j}^{k+1} + \alpha n|_{i,j}^{k+1} - \theta u|_{i,j}^{k+1} m|_{i,j}^{k+1} - \beta m|_{i,j}^{k+1}),$$

$$u|_{i,j}^{k+1} = u|_{i,j}^{k+1/2} + \frac{\Delta t}{2}(d_u \Delta u|_{i,j}^{k+1} + \xi f|_{i,j}^{k+1} - \theta u|_{i,j}^{k+1} m|_{i,j}^{k+1} - \rho u|_{i,j}^{k+1}).$$

To solve the systems of equations, the Gauss-Seidel method was used. More details can be see in Maganin (2020).

## 4 NUMERICAL SIMULATIONS

The numerical simulations for the equation model (1)-(4) are performed in the rectangular domain $[x_0, x_f] \times [y_0, y_f]$. To solve the system numerically, the dimensionless partial differential equations are considered, as described in Anderson *et al*. (2000), Capelão and Lolas (2005), Chaplain, Anderson and Preziosi (2003) and Ganesan and Lingeshwaran (2017).

### 4.1 Verification using the Method of Manufactured Solutions

According to Vargas (2013), due to the existence of analytical solutions only for models involving simpler equations, the main difficulty in evaluating the discretization error is to find a way to estimate the analytical solution for EDP's and thus obtain greater reliability in the analyzes.

One technique that can help to verify the implementation of the computational code, through the convergence analysis by refinement, is the Method of Manufactured Solutions (MMS).

Thus, a study of the convergence of (1)-(4) is presented, using the norm of the relative error $L_2$, found in Lima (2010). Exact solutions are considered as reference solutions as presented in Ganesan and Lingeshwaran (2017). So, we have the tumor growth model, coupled with the source terms

$$\frac{\partial n}{\partial t} - d_n \nabla^2 n + \gamma \nabla \cdot (n \nabla f) - \mu_1 n(1 - n - f) = f_n, \tag{5}$$

$$\frac{\partial f}{\partial t} + \eta m f - \mu_2 f(1 - n - f) = f_f, \tag{6}$$

$$\frac{\partial m}{\partial t} - d_m \nabla^2 m - \alpha n + \theta u m + \beta m = f_m, \tag{7}$$

$$\frac{\partial u}{\partial t} - d_u \nabla^2 u - \xi f + \theta u m + \rho u = f_u. \tag{8}$$

The source terms $f_n$, $f_f$, $f_m$ and $f_u$ are obtained so that the equations (5)-(8) satisfy the analytical solutions $n(x, y, t) = m(x, y, t) = u(x, y, t) e^t \sin(2\pi x) \sin(2\pi y)$ and $f(x, y, t) = e^{-t} \sin(2\pi x) \sin(2\pi y)$. This procedure, characterizes the process of manufactured solutions.

For the analysis of the convergence study, a domain is considered $\Omega = [0.1] \times [0.1]$, with final time $t_f = 0.25$ and dimensionless parameters $d_n = 0.001$, $d_m = 0.001$ and $d_u = 0.001$, $\gamma = 0.01$, $\mu_1 = 0.5$, $\eta = 10$, $\mu_2 = 0.1$, $\alpha = 0.1$, $\beta = 0.07$, $\theta = 0.05$, $\xi = 0.03$ and $\rho = 0.07$, found in Lopez, Ruiz and Castaño (2018). Initial and boundary conditions are obtained using the analytical solutions of $n, f, m$ and $u$. The results obtained, using the norm of the relative error $L_2$, are presented in Table 1, where different refinements are evaluated, both in time and space.

Table 1 - Relative error of the norm $L_2$ for $t_f = 0.25$.

| $M_t$ | Erro $n$ | Erro $f$ | Erro $m$ | Erro $u$ |
|---|---|---|---|---|
| $M_x = M_y = 20$ | | | | |
| 20 | 2.5713x10$^{-2}$ | 5.1190x10$^{-4}$ | 3.9675x10$^{-4}$ | 8.2419x10$^{-5}$ |
| 40 | 2.4724x10$^{-2}$ | 3.6709x10$^{-4}$ | 3.3785x10$^{-4}$ | 3.8769x10$^{-5}$ |
| 80 | 2.4245x10$^{-2}$ | 2.8345x10$^{-4}$ | 3.0949x10$^{-4}$ | 1.7887x10$^{-5}$ |
| 160 | 2.4008x10$^{-2}$ | 2.3901x10$^{-4}$ | 2.9557x10$^{-4}$ | 7.6797x10$^{-6}$ |
| 320 | 2.3891x10$^{-2}$ | 2.1616x10$^{-4}$ | 2.8869x10$^{-4}$ | 2.6343x10$^{-6}$ |
| $M_x = M_y = 40$ | | | | |
| 20 | 1.7346x10$^{-2}$ | 2.4236x10$^{-4}$ | 2.7762x10$^{-4}$ | 7.3674x10$^{-5}$ |
| 40 | 1.6492x10$^{-2}$ | 1.4556x10$^{-4}$ | 2.2726x10$^{-4}$ | 3.5525x10$^{-5}$ |
| 80 | 1.6085x10$^{-2}$ | 8.9821x10$^{-5}$ | 2.0307x10$^{-4}$ | 1.7255x10$^{-5}$ |
| 160 | 1.5886x10$^{-2}$ | 6.0241x10$^{-5}$ | 1.9122x10$^{-4}$ | 8.3198x10$^{-6}$ |
| 320 | 1.5788x10$^{-2}$ | 4.5038x10$^{-5}$ | 1.8536x10$^{-4}$ | 3.9019x10$^{-6}$ |
| $M_x = M_y = 80$ | | | | |
| 20 | 9.9788x10$^{-3}$ | 1.5260x10$^{-4}$ | 1.8815x10$^{-4}$ | 8.0794x10$^{-5}$ |
| 40 | 9.0420x10$^{-3}$ | 8.5080x10$^{-5}$ | 1.3525x10$^{-4}$ | 3.9212x10$^{-5}$ |
| 80 | 8.6061x10$^{-3}$ | 4.6238x10$^{-5}$ | 1.0992x10$^{-4}$ | 1.9265x10$^{-5}$ |
| 160 | 8.3962x10$^{-3}$ | 2.5632x10$^{-5}$ | 9.7546x10$^{-5}$ | 9.5023x10$^{-6}$ |
| 320 | 8.2932x10$^{-3}$ | 1.5043x10$^{-5}$ | 9.1427x10$^{-5}$ | 4.6732x10$^{-6}$ |

It is observed, in Table 1, that the error order of *n, f, m* and *u* decreased when refining the mesh, both spatial and temporal, where was considered $M_t = 2^n \times 20, n = 0,...,4$ , $M_x = M_y = 2^m \times 20, m = 0,1,2$ and $t_f = 0.25$.

Still, the results of the table show that the order of the relative error, norm $L_2$, when considering the refinements $M_t = 160$ and $M_t = 360$ remained the same, in all the analyzed variables, not requiring a high refinement and consequently reducing computational costs. Also, on spatial refinement, $M_x = M_y$ from 40 to 80, it is observed that for *u*, there was a small increase, but maintained the same order of convergence. Finally, through the results obtained, it can be said that the two-stage method and the linearization of the quadratic terms used in the model (1)-(4), are satisfactory when compared with analytical solutions through error analysis. Therefore, it appears that the implementation of the code, through the manufactured solution, proved to be efficient.

**4.2 Breast tumor simulations**

The objective of this work is in the simulation and analysis of the tumor growth model in a non-regular mesh, which represents the geometry of a female breast, simulating a problem close to reality.

The breast geometry to be used is illustrated in Foucher, Ibrahim and Saad (2018). This geometry is used to present an analysis of tumor development in the model proposed in this work.

The points of the contour of the breast geometry were collected using the program *WebPlotDigitizer* 4.3 (ROHATGI, 2020). The contour is illustrated in Figure1. Mesh refinement $M_x = M_y = 100$ is considered for a breast approximately 14 cm wide by 18 cm high. It is observed that the points of the geometry in the first column of Figure 1, given outline, blue color, are not part of the mesh, therefore, the technique that consists of approximating the contour of the domain by mesh segments is employed, seen in Cuminato and Meneguette (2013).

The comparison of the approximate contour, red color, and given, blue color, is found in the second column of Figure 1, it can be seen better with the zoom. In the third column, the approximate outline is shown with the internal points of the mesh, green color, for simulation.

To approximate the domain contour by mesh segments, a code was developed that locates the points closest to the vertex in the mesh about the given contour point, resulting in coordinates of the vertices in the mesh scale, generating the approximate contour. To obtain the approximate contour, the code was divided into parts, where:

i) Determine in which part of the figure the point is located, right, left, above or below the given contour point;
ii) All points near the approximate contour are located;
iii) There is a variation between the approximate contour point and the given one,
iv) Points are calculated between two vertices with variations in the directions *x* and *y*.

Thus, the model starts to be solved considering the new domain, in which it is assumed that the boundary conditions apply under the new boundary. In this way, the model (1)-(4) will be solved at the internal points of the geometry, illustrated by the green color in the last column of Figure 1.

Figure 1: Contour given and approximate of a breast.

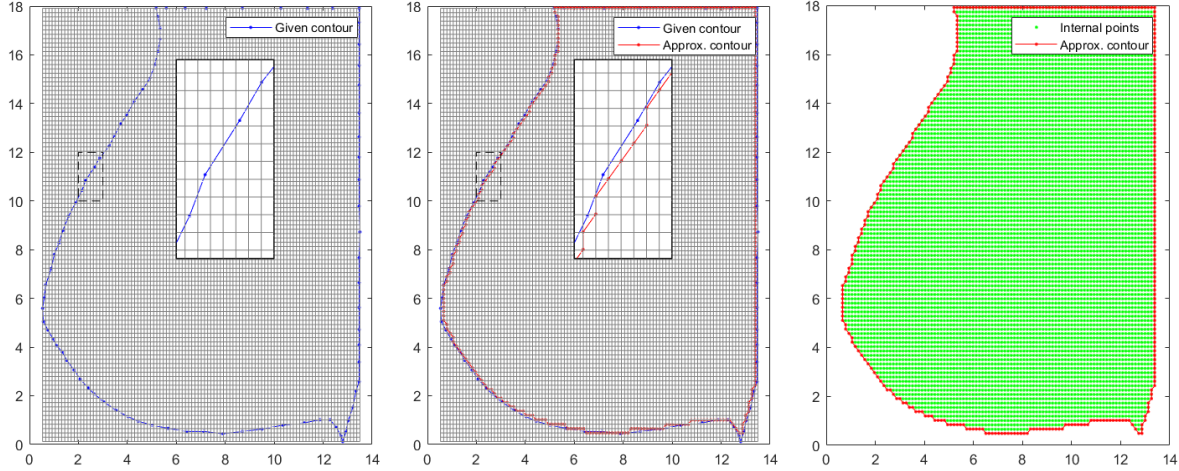

To solve the model within the breast geometry, the initial conditions described in (9)-(12), where it is initially assumed that there is a nodule of cells present where the initial tumor density in two dimensions is at point (6.88, 4.39), that is, the tumor has degraded some of its surrounding tissues and, therefore, the initial conditions are given by Chaplain, Anderson and Preziosi (2003) and by Kolev and Zubik-Kowal (2011),

$$n(x, y, 0) = exp\left(-\frac{r^2}{\epsilon}\right), \quad (9)$$

$$f(x, y, 0) = 1 - 0.5\, n(x, y, 0), \quad (10)$$

$$m(x, y, 0) = 0.5\, n(x, y, 0), \quad (11)$$

$$u(x, y, 0) = 0, \quad (12)$$

being $(x, y) \in \Omega$ and $r$ is defined by $r = \sqrt{(x - 6.88)^2 + (y - 4.39)^2}$.

The model (1)-(4) is simulated at the internal points of the breast geometry, with the dimensionless parameters mentioned, seen in Lopez, Ruiz and Castaño (2018), with $\epsilon = 0.07$. The choice of the value of $\epsilon = 0.07$ describes an initial tumor with 1 centimeter in diameter at $t = 0$, because according to INCA (2019), it is the minimum tumor size that the self- breast exam or clinical examination, performed by the doctor can detect. In the simulations, the tumor growth is evaluated up to 5 centimeters, since from this size the tumor may present metastases and, the mathematical model does not describe tumors in the vascular stage.

Figure 2 shows the initial tumor condition of the variables $n$, $f$, $m$ and $u$, in a, b, c and d, respectively, centered at approximately (6.88, 4.39). The tumor in the breast geometry is approximately 1 centimeter in diameter, in the mesh with $M_x = M_y = 100$ and $M_t = 1280$.

As for the boundary conditions, only Dirichlet-type conditions are considered, in such a way that it has null values for $n$, $m$ and $u$ The values of the points in the contour for $f$ were considered equal to 1.

It is observed in Figure 2, that the initial conditions of the model variables, present the presence of tumor cells, resulting in a loss of extra-cellular matrix density and the appearance of matrix degradative enzymes. Only the variable $u(x, y, t)$, which represents endogenous inhibitors, does not contain concentration when $t = 0$.

Figure 2: Initial condition of variables $n$, $f$, $m$ and $u$.

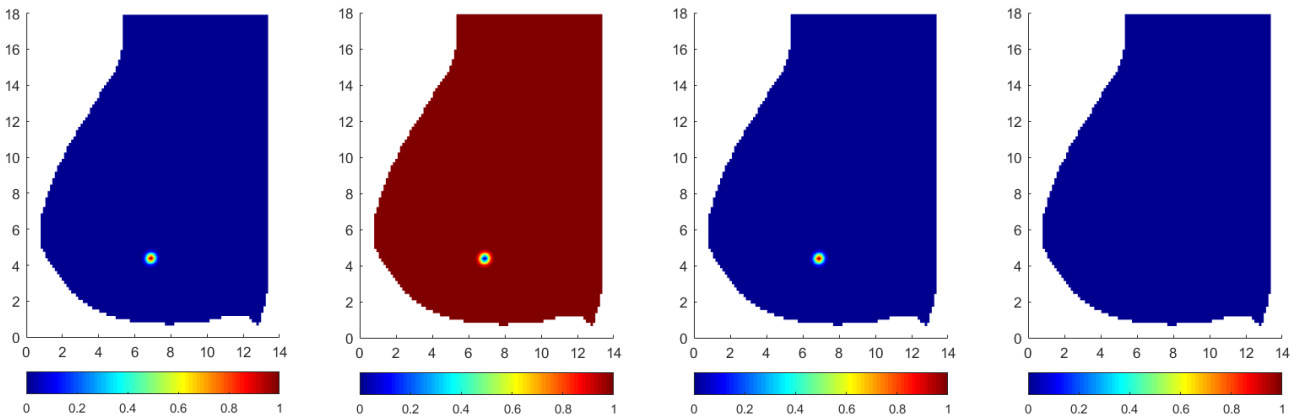

To analyze the growth of the tumor, it is found in Lee *et al.* (2016), a study that evaluated the tumor growth rate of breast cancer. Through this study, the authors presented the rate of tumor growth, per day, where tumors with aggressive molecular subtypes have a growth rate of $1.003 % d$^{-1}$. This rate will be considered in this work, to analyze the tumor growth model, in days, in the breast geometry. It is observed that with the growth rate of 1.003 % per day, the tumor reaches 5 cm in diameter in approximately 161 days.

Figure 3 shows the evolution of the growth of tumor cells n, consequently the loss of density of the extracellular matrix f, MED m and endogenous inhibitors u for time $t = 40$, that using the tumor growth rate equivalent to 1.003 %, corresponds approximately to 161 days.

Figure 3: Behavior of the variables $n$, $f$, $m$ and $u$ in $t = 40$.

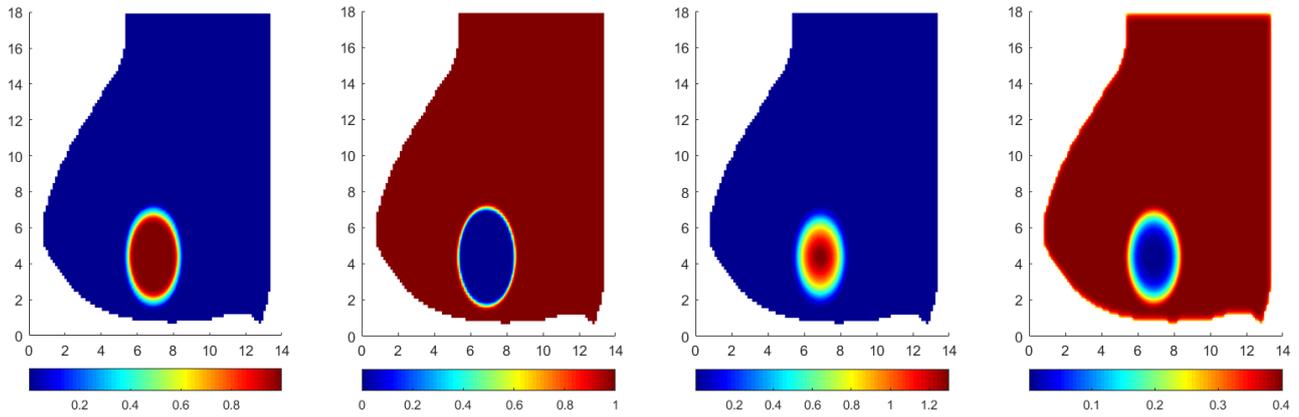

It can be observed, Figure 3, that the model describes the interactions predicted in the modeling, that is, the evolution of the growth of the tumor cells implies the loss of the density of the extracellular matrix.

As a second test, the analysis for the model is presented, in which it is assumed that the tumor was detected near the outline of the breast geometry, centred on (2.2208; 8.8396), Figure 4.

Figure 4: Initial condition of variables $n$, $f$, $m$ and $u$ in position closest to the contour.

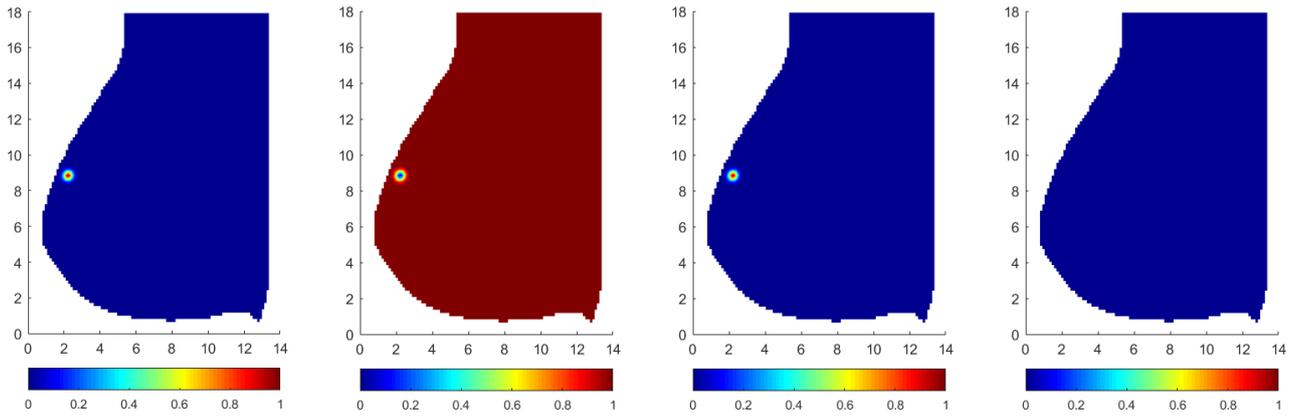

In Figure 5, we have the results of the evolution of the four variables analyzed, in the position close to the breast contour, in $t = 40$.

Figure 5: Behavior of the variables $n$, $f$, $m$ and $u$ in position closest to the contour

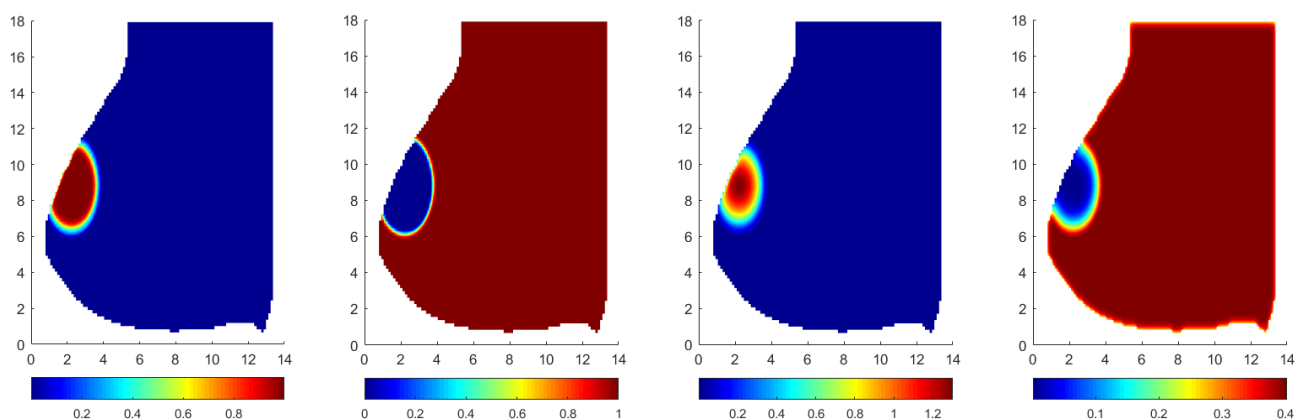

Considering the position of the tumor closest to the left contour, it became possible to verify how the distribution of the density of cancer cells, the density of MEC, concentration of EDM and TIMP is found. It is observed that with the growth of these variables, for $t = 40$, the circular shape is lost since the boundary condition is null at the border and there is no presence of tumor cells outside the breast region.

## 5 CONCLUSIONS

The purpose of this work was to conduct a study of the mathematical model of tumor growth by Kolev and Zubik-Kowal (2011). The simulations showed that the model has important characteristics of the interactions between tumor cells and the surrounding tissue, since, as there is an increase in cancer cells, the loss of extracellular matrix occurs, which is responsible for cell adhesion, cell-a communication -cell and differentiation. In addition, the simulations present the concentration of degradative enzymes, which has the role of assisting tumor cells in the process of degradation of the ECM.

## 6 ACKNNOWLEDGMENTS

This study was financed in part by the Coordenação de Aperfeiçoamento de Pessoal de Nível Superior - Brasil (CAPES) - Finance Code 001.